\documentclass[aps,prl,showpacs,twocolumn,
nofootinbib]{revtex4}
\usepackage[dvips]{epsfig}
\usepackage{graphicx}
\usepackage{amsmath}

\begin{document}

\title{{\Large \textbf{Photon as a Vector Goldstone Boson: Nonlinear $\sigma $ \
Model for QED}}}
\author{\textbf{J.L.~Chkareuli$^a$}, \textbf{C.D.~Froggatt$^b$}, \textbf{R.N.
Mohapatra$^c$} and \textbf{H.B. Nielsen$^d$}}

\affiliation{$^a$ Institute of Physics, Georgian Academy of Sciences,
0177 Tbilisi, Georgia}

\affiliation{ $^b$Department of Physics and Astronomy, Glasgow
University}

\affiliation{$^c$Department of Physics, University of Maryland, College
Park, MD-020742, USA}

\affiliation{$^d$ Niels Bohr Institute, Blegdamsvej 17-21, DK 2100
Copenhagen, Denmark}

\date{December, 2004}

\begin{abstract}
We show that QED in the Coulomb gauge can be considered as a low energy
linear approximation of a non-linear $\sigma $-type model where the photon
emerges as a vector Goldstone boson related to the spontaneous breakdown of
Lorentz symmetry down to its spatial rotation subgroup at some high scale $M$%
. Starting with a general massive vector field theory one naturally arrives
at this model if the pure spin-1 value for the vector field $A_{\mu }(x)$
provided by the Lorentz condition $\partial _{\mu }A_{\mu }(x)=0$ is
required. The model coincides with conventional QED in the Coulomb gauge for 
$M\rightarrow \infty$ and generates a very particular form for the Lorentz
and CPT symmetry breaking terms, which are suppressed by powers of $M$.
\end{abstract}

\maketitle

%%%%%%%%%%%%%%%%%%%%%%%%%%%%%%%%%%%%%%%%%%%%%%%%%%%%%%%
\vskip1.0in

\section{Introduction}

\label{introduction}In recent years there has been considerable interest in
the breakdown of the Lorentz invariance, as a phenomenological possibility 
\cite{alan} in the context of various quantum field theories as well as
modified gravity and string theories\cite{kraus,arkani,grip,tajac}. The
existence of Lorentz violation leads to a plethora of new high energy
effects \cite{alan,glashow} with interesting implications for neutrino
experiments as well as high energy cosmic ray phenomena. Furthermore, for
particular parameterizations of Lorentz violation, these high energy effects
can also lead to bounds on the strengths of these effects.

In this note we discuss Lorentz violating effects within a new formulation
of Quantum Electrodynamics (QED), where the familiar QED in Coulomb gauge
emerges as an effective low energy Lagrangian in a theory where the
masslessness of photon is related to spontaneous breakdown of the Lorentz
symmetry (SBLS). Lorentz symmetry and its spontaneous breakdown is an old
idea\cite{bjorken} and have been considered in many different contexts\cite
{book}, particularly, in the generation of the internal symmetries observed
in particle physics, although many formulations of the idea look
contradictory (see for some recent criticism \cite{kraus,jenkins}). Our
point is that an adequate formulation of the SBLS should be related to a
fundamental vector field Lagrangian by itself rather than an effective field
theory framework containing a (finite or infinite) set of the primary
fermion interactions where the vector fields appear as the auxiliary fields
for each of these fermion bilinears\footnote{%
There is no need, in essence, for the physical SBLS to generate ''composite
'' vector bosons (related with fermion bilinears) which could mediate the
gauge-type binding interactions in Abelian or non-Abelian theories\cite
{bjorken,book}.This conclusion most clearly follows from the proper lattice
formulation \cite{randjbar} where Lorentz invariance is explicitly broken at
the very beginning.}.

Actually, the symmetry structure of the existing theories, like as QED or
Yang-Mills Lagrangians, seems to be well consistent with such a point of
view \cite{cfn}. The vector field gauge-type transformation, say for QED, of
the form 
\begin{equation}
A_{\mu }(x)\rightarrow A_{\mu }(x)+n_{\mu }\text{ \ \ \ \ \ \ }(\mu
=0,1,2,3),  \label{vector}
\end{equation}
can be identified by itself as the pure SBLS transformation with vector
field $A_{\mu }(x)$ developing some constant background value $n_{\mu }$%
\footnote{%
Remarkably, it was argued a long time ago \cite{picasso} that just the
invariance of the QED under the transformations (\ref{vector}) with a gauge
function linear in the co-ordinates ($A_{\mu }\rightarrow A_{\mu }+\partial
_{\mu }\omega $, $\omega =n_{\mu }x^{\mu }$) implies that the theory
contains a genuine zero mass vector particle.}. The point is, however, this
Lorentz symmetry breaking does not manifest itself in any physical way, due
to the fact that an invariance of the QED under the transformation (\ref
{vector}) leads to the conversion of the SBLS into gauge degrees of freedom
of the massless photon. This in essence is what we call the
non-observability of the SBLS of type (\ref{vector}). In this connection it
was recently shown \cite{cfn} that gauge theories, both Abelian and
non-Abelian, can be obtained from the requirement of the physical
non-observability of the SBLS (\ref{vector}), caused by the Goldstonic
nature of vector fields, rather than from the standard gauge principle.

It is instructive here to compare this QED case with the free massless
(pseudo-) scalar triplet theory ($\partial ^{2}\phi ^{i}=0$), which is
invariant under a similar spontaneous symmetry breaking transformation 
\begin{equation}
\phi ^{i}(x)\rightarrow \phi ^{i}(x)+c^{i}\qquad (i=1,2,3),  \label{scalar}
\end{equation}
where the $c^{i}$ are arbitrary constants. Again this symmetry
transformation corresponds to zero-mass excitations of the vacuum, which
might be identified (in some approximation) with physical pions. However, in
marked contrast with the vector field case (\ref{vector}) where
renormalizable interactions of the form $\bar{\psi}\gamma _{\mu }\psi A^{\mu
}$ are invariant under the transformation (\ref{vector}) (accompanied by $%
\psi \rightarrow e^{in\cdot x}\psi $), the scalar field theory invariant
under (\ref{scalar}) ends up being a trivial theory. The only way to have it
as an interacting theory is to add additional states such as the $\sigma $
field as in the famous $\sigma $-model\cite{GMlevy} of Gell-Mann and Levy.

A question then is whether one can have an alternative formulation of QED
where the symmetry transformation in (\ref{vector}) is realized in a manner
analogous to the $\sigma $-model and, if so, what kind of new physics it
leads to in addition to the familiar successes of QED.

It is quite clear that the simplest way to retain the explicitly covariant
form of the vector Goldstone boson transformation (\ref{vector}), is to
enlarge the existing Minkowskian space-time to higher dimensions with our
physical world assumed to be located on a three-dimensional Brane embedded
in the high-dimensional bulk. However, while technically it is quite
possible to start, say, with the spontaneous breakdown of the 5-dimensional
Lorentz symmetry $SO(1,4)\rightarrow SO(1,3)$ \cite{Li} to generate an
ordinary 4-dimensional vector Goldstone vector field $A_{\mu }(x)$, a
serious problem for such theories is how to achieve the localization of this
field on the flat Brane associated with our world \cite{rubakov}. We
therefore take an alternative path: we start with a general massive vector
field theory in an ordinary 4-dimensional space-time. The only restriction
imposed is the requirement that the four-vector $A_{\mu }$ in order to
describe a spin-1 particle, must satisfy the Lorentz condition\footnote{%
This supplementary condition is in fact imposed as an off-shell constraint,
independent of its equation of motion\cite{ogi}.} 
\begin{equation}
\partial _{\mu }A^{\mu }(x)=0  \label{spin}
\end{equation}
In this connection, it seems important to note that we are dealing further
with just the physical vector field condensation rather than a condensation
of the scalar component in the 4-vector $A_{\mu }(x)$, as might occur in the
general case when the supplementary Lorentz condition (\ref{spin}) is not
imposed. We show that this leads to a non-linear $\sigma $-type model for
QED, where the photon emerges as a vector Goldstone boson related to the
spontaneous breakdown of Lorentz symmetry down to its spatial rotation
subgroup $SO(1,3)\rightarrow SO(3)$ at some high scale $M$. The model
appears to coincide with ordinary QED taken in the Coulomb gauge in the
limit where the scale $M$ goes to infinity. For finite values of $M$, there
appear an infinite number of nonlinear photon interaction and
self-interaction terms properly suppressed by powers of $M$. These terms
violate Lorentz invariance and could have interesting implications for
physics.

\section{The Spin-1 Vector Fields and Physical SBLS}

\label{abelian} Let us consider a simple Lagrangian for the neutral vector
field $A_{\mu }(x)$ and one fermion $\psi (x)$ with dimensionless coupling
constants ( $\lambda $ and $e$) 
\begin{equation}
\mathcal{L}=-\frac{1}{4}F_{\mu \nu }F^{\mu \nu }+\frac{\mu ^{2}}{2}A_{\mu
}^{2}-\frac{\lambda }{4}(A_{\mu }^{2})^{2}+\overline{\psi }(i\gamma \partial
-m)\psi -eA_{\mu }\overline{\psi }\gamma ^{\mu }\psi   \label{Lagr}
\end{equation}
where $F_{\mu \nu }=\partial _{\mu }A_{\nu }-\partial _{\nu }A_{\mu }$
denotes the field strength tensor for the vector field $A_{\mu
}=(A_{0},A_{i})$ (and we denote $A_{\mu }A^{\mu }\equiv A_{\mu }^{2}$ and
use similar shorthand notation e.g. $(\partial _{\mu }A_{i})^{2}\equiv
\partial _{\mu }A_{i}\partial ^{\mu }A_{i}$ etc. later). The free part of
the Lagrangian is taken in the standard form so that in this case the
Lorentz condition (\ref{spin}) automatically follows from the equation for
the vector field $A_{\mu }$. The $\lambda A^{4}$ term is added to implement
the spontaneous breakdown of Lorentz symmetry $SO(1,3)$ down to the $SO(3)$
or $SO(1,2)$ for $\mu ^{2}>0$ and $\mu ^{2}<0$ , respectively. Note that
contribution of the form $A^{4}$ (and higher)\footnote{%
In fact, one might add one more term of the form $A_{\mu }A^{\nu }\partial
_{\nu }A^{\mu }$, making the neutral vector field Lagrangian (\ref{Lagr}) to
be the most general parity-conserving theory with only terms of dimension $%
\leq 4$. However, in the ground state of interest here, this extra term
vanishes and leads to the same physics (see below).} naturally arises in
string theories\cite{alan,alan1}.

Writing down the equation of motion for vector and fermion fields 
\begin{eqnarray}
\partial ^{2}A_{\mu }-\partial _{\mu }\partial ^{\nu }A_{\nu }+\mu
^{2}A_{\mu }-\lambda A_{\mu }A_{\nu }^{2}-e\overline{\psi }\gamma _{\mu
}\psi  &=&0  \label{vec} \\
(i\gamma \partial -m)\psi -eA^{\mu }\gamma _{\mu }\psi  &=&0  \label{fer}
\end{eqnarray}
taking then the 4-divergence of Eq.(\ref{vec}), % replacing there $\square
and requiring that the Lorentz condition (\ref{spin}) be fulfilled, one
comes to the equation 
\begin{equation}
\lambda \partial _{\nu }(A^{\mu }A_{\mu })~=0  \label{cond}
\end{equation}
which should be satisfied identically. Otherwise, it would represent by
itself one more supplementary condition (in addition to the equations of
motions and Lorentz condition) implying that the field $A_{\mu }$ has fewer
degrees of freedom than is needed for describing all its three possible spin
states. % (provided that it is left massive if the equation (\ref
%{cond}) is taken to work).
This is definitely inadmissible. Furthermore a solution of the form $\lambda
=0$ is also not acceptable for $\mu ^{2}<0$ since in this case, the
Hamiltonian for the theory has no lower bound. Thus the only solution to Eq.
(\ref{cond}) for the physical massive vector field corresponds generally to
the case 
\begin{equation}
\lambda \neq 0,\text{ \ \ \ \ }A_{\mu }^{2}=M^{2}
\end{equation}
where $M^{2}=\frac{\mu ^{2}}{\lambda }$ stands for some arbitrary constant
parameter with dimensionality of mass squared. The general Lagrangian (\ref
{Lagr}) now takes the form 
\begin{equation}
\mathcal{L}_{SBLS}=-\frac{1}{4}F^{\mu \nu }F_{\mu \nu }+\overline{\psi }%
(i\gamma \partial +m)\psi -eA^{\mu }\overline{\psi }\gamma _{\mu }\psi
+const,  \label{lagr1}
\end{equation}
with the important constraint 
\begin{equation*}
A_{0}^{2}-A_{i}^{2}=M^{2},
\end{equation*}
from which it will be obviously noticed that the vector field $A_{\mu }$
appears massless, while its vev leads to the actual SBLS. We have obtained
in fact the nonlinear $\sigma $-type model for QED which we now expand in
more detail. Remarkably, there is no other solution to our basic equation (%
\ref{cond}) inspired solely by the spin-1 requirement for massive vector
field (\ref{spin}). Furthermore, the condition (8) arises regardless of the
sign of $\mu ^{2}$ leading to the global minimum of the theory (given by
constant term in $\mathcal{L}_{SBLS}$) which locates lower than the case
where the vector field has zero vev. However, for the right sign mass term
case ($\mu ^{2}>0$) taken in  the starting Lagrangian (\ref{Lagr}) the
Lorentz symmetry always breaks down to its spatial rotation subgroup $%
SO(1,3)\rightarrow SO(3)$. This is the main result of the paper, whose
implications we study now\footnote{%
Note that some models of QED with a nonlinear condition $%
A_{0}^{2}-A_{i}^{2}=M^{2}$ has been previously considered \cite
{nambu,venturi,alan1}. In refs. \cite{nambu,venturi}, this condition was
used as a pure gauge choice whereas in our case it is a dynamical constraint
(stemming from the spin-1 requirement for vector field) since our basic
massive vector field Lagrangian does not have gauge invariance. In Ref.\cite
{alan1}, this condition appeared as a symmetry breaking condition in the
string-inspired models (called ``bumblebee models'') with an effective
negative-sign mass square term for vector field. A crucial difference 
between the
present work and the work of ref. \cite{alan1} is that in our case the
dynamical spin-1 requirement (\ref{spin}) leads to the constraint in Eq.(8)
for vector field regardless of the sign of its mass-term. Apart from the
fact that for the case of right (positive) sign for $\mu ^{2}$ one has the
global minimum of the theory for the SBLS of type $SO(1,3)\rightarrow SO(3)$
, this requirement also excludes any ghost like modes in the model. As a
result, our model is fundamentally different, with a very different
effective Lagrangian at low energies.}.

\section{ Nonlinear $\protect\sigma $ Model for QED}

The above considerations allow us to argue that the spin-1 vector field $%
A_{\mu }$ can selfconsistently be presented in the Lorentz symmetry phase as
the massive vector field mediating the fermion (and any other matter)
interactions in the framework of the massive QED, or, conversely, in the
physical SBLS phase (\ref{lagr1}) as the basic condensed field producing
massless Goldstone states which then are identified with physical photons.

Taking the characteristic SBLS parameter $M^{2}$ positive ($M^{2}>0$) one
comes to the breakdown of the Lorentz symmetry to its spatial rotation
subgroup with the vector field space-components $A_{i}$ ($i=1,2,3$) as the
Goldstone fields. Their Lagrangian immediately follows from Eq.(\ref{lagr1})
which after using the Lorentz condition (\ref{spin}) and elimination of the
vector field time-component $A_{0}$ looks like 
\begin{eqnarray}
\mathcal{L}_{QED\sigma } &=&-\frac{1}{2}(\partial _{\mu }A_{\nu })^{2}+%
\overline{\psi }(i\gamma \partial +m)\psi -eA^{\mu }\overline{\psi }\gamma
_{\mu }\psi  \notag \\
&=&\frac{1}{2}(\partial _{\mu }A_{i})^{2}-\frac{1}{2}\frac{(A_{i}\partial
_{\mu }A_{i})^{2}}{M^{2}+A_{i}^{2}}+  \label{lagr2} \\
&&+\overline{\psi }(i\gamma \partial +m)\psi +eA_{i}\overline{\psi }\gamma
_{i}\psi  \notag \\
&&-e\sqrt{M^{2}+A_{i}^{2}}\overline{\psi }\gamma _{0}\psi  \notag
\end{eqnarray}
We now expand the newly appeared terms in powers of $\frac{A_{i}^{2}}{M^{2}}$
and also make the appropriate redefinition of fermion field $\psi $
according to 
\begin{equation}
\psi \rightarrow e^{ieMx_{0}}\psi
\end{equation}
so that the mass-type term $eM\overline{\psi }\gamma _{0}\psi $, appearing
from the expansion of the fermion current time-component interaction in the
Lagrangian (\ref{lagr2}) will be exactly cancelled by an analogous term
stemming now from the fermion kinetic term. After this redefinition, and
collecting the linear and nonlinear (in the $A_{i}$ fields) terms
separately, we arrive at the Lagrangian 
\begin{eqnarray}
\mathcal{L}_{QED\sigma } &=&\frac{1}{2}(\partial _{\mu }A_{i})^{2}+\overline{%
\psi }(i\gamma \partial +m)\psi +eA_{i}\overline{\psi }\gamma _{i}\psi - \\
&&-\frac{1}{2}\frac{(A_{i}\partial _{\mu }A_{i})^{2}}{M^{2}}\left( 1-\frac{%
A_{i}^{2}}{M^{2}}+\cdot \cdot \right)  \notag \\
&&-e\frac{A_{i}^{2}}{2M}\left( 1-\frac{A_{i}^{2}}{4M^{2}}\cdot \cdot \cdot
\right) \overline{\psi }\gamma _{0}\psi  \notag
\end{eqnarray}
where we have retained the former notation for the fermion $\psi $ and
omitted the higher nonlinear terms for photon. Additionally, the Lorentz
condition for the spin-1 vector field (\ref{spin}) now reads as follows: 
\begin{equation}
\partial _{i}A_{i}-\frac{A_{i}\partial _{0}A_{i}}{M}\left( 1-\frac{A_{i}^{2}%
}{2M^{2}}+\cdot \cdot \cdot \right) =0  \label{spin2}
\end{equation}

The Lagrangian (12) together with a modified Lorentz condition (\ref{spin2})
completes the $\sigma $ model construction for quantum electrodynamics. We
will call this $QED\sigma$. The model contains only two independent (and
approximately transverse) vector Goldstone boson modes which are identified
with the physical photon, and in the limit $M\rightarrow \infty $ is
indistinguishable from conventional QED taken in the Coulomb gauge. In this
limit the 3-dimensional analog of the ``goldstonic'' gauge transformations (%
\ref{vector}) accompanied by the proper phase transformation of fermion 
\begin{equation}
A_{i}(x)\rightarrow A_{i}(x)+n_{i},\ \ \ \ \ \ \ \psi \rightarrow
e^{-ien_{i}x_{i}}\psi \ \ \ \ \ \ \ (i=1,2,3)
\end{equation}
emerges as an exact symmetry of the Lagrangian in Eq. (12), as one would
expect in the pure Goldstonic phase.

While $QED\sigma $ coincides with the conventional QED in Coulomb gauge in
the limit of $M\rightarrow \infty $, it differs from the conventional QED in
Coulomb gauge in several ways. First, apart from an ordinary photon-fermion
coupling, our model generically includes an infinite number of nonlinear
photon interaction and self-interaction terms which become active at high
energies comparable to the SBLS scale $M$. Second and more important, the
nonlinear photon interaction terms in the Lagrangian Eq. (12) break Lorentz
invariance in a very specific way depending only on a single parameter $M$
unlike many recent parameterizations of Lorentz breaking which involve more
than one new parameter. Furthermore all the non-linear photon-fermion
(photon-matter in general) interaction terms are C, CP and CPT non-invariant
as well. This should have interesting implications for particle physics and
cosmology, such as high precision measurements involving atomic systems,
breaking of C and CP invariance in electromagnetic processes, extra
contribution to neutral mesons oscillations and, especially, the
implications of CPT-violating effects on the matter-antimatter asymmetry in
the early universe. We will pursue these implication in a separate
publication. One immediate point to note is that the dispersion formula for
light propagation still remains the same (i.e. $\omega _{k}^{2}-|\vec{k}%
|^{2}~=~0$).

\section{ Conclusion}

To summarize, we have started with the observation\cite{cfn} that the
gauge-type transformations (\ref{vector}) with a gauge function linear in
the co-ordinates can be treated as the transformations of the spontaneously
broken Lorentz symmetry, whose pure Goldstonic phase is presumably realized
in the form of the known QED. Exploring this point of view and starting from
the general massive vector field theory, we have constructed a full
theoretical framework for the physical SBLS including its Higgs phase as
well, in terms of the properly formulated nonlinear $\sigma $-type model .
For the first time we have proposed a pure fundamental Lagrangian
formulation without referring to the effective four-fermion interaction
ansatz dating back to the pioneering work of Bjorken \cite{bjorken}.

In this connection, one might conclude that the whole non-linear Lagrangian $%
\mathcal{L}_{QED\sigma }$ (12), with a massless photon provided by the
spontaneous breakdown of Lorentz invariance, is in some sense a more
fundamental theory of electromagnetic interactions than the usual QED%
\footnote{%
Indeed this origin for the masslessness of the photon seems to be more
general and deep than the usually postulated gauge symmetry. Despite the
essentially non-renormalisable character of the Lagrangian $\mathcal{L}%
_{QED\sigma }$ (12), one does not expect the radiative corrections to
generate a mass for the photon; otherwise one would have to admit that the
radiative corrections lead to a breakdown of the original Lorentz symmetry
in the starting Lagrangian (\ref{Lagr}) or (\ref{lagr1}), which is hardly
imaginable.}. This theory coinciding with quantum electrodynamics at low
energies happens to generically predict striking new phenomena beyond
conventional QED at high energies comparable to the SBLS scale $M$ - an
infinite number of nonlinear photon-photon and photon-matter interactions
which explicitly break relativistic invariance, and C, CP and CPT symmetry.

\section*{Acknowledgments}

We would like to thank Gia Dvali, Oleg Kancheli, Alan Kostelecky, Gordon
Moorhouse, Valery Rubakov, David Sutherland and Ching Hung Woo for useful
discussions and comments. One of us (J.L.C.) is grateful for the kind
hospitality shown to him during his summer visit (June-July 2003) to the
Department of Physics and Astronomy at Glasgow University, where part of
this work was carried out. Financial support from GRDF grant No. 3305 is
also gratefully acknowledged by J.L.C. and R.N.M. R. N. M. is supported by
National Science Foundation Grant No. PHY-0354401.

\end{document}